\documentclass[sigconf]{acmart}
\AtBeginDocument{%
  }

\copyrightyear{2025}
\acmYear{2025}
\setcopyright{cc}
\setcctype{by-nc-sa}
\acmConference{Access InContext Workshop @ CHI’25}{April 26}{Yokohama, Japan}
\acmBooktitle{Access InContext Workshop @ CHI’25, April 26, 2025, Yokohama, Japan}

\begin{document}

\title{Subtitled Media Adaptations for People with Aphasia}
\subtitle{Ongoing Accessibility Barriers and Emerging Design Practices}

\author{Zihao You}
\orcid{0000-0002-3450-0543}
\affiliation{%
  \institution{School of Science and Engineering\\University of Dundee}
  \city{Dundee}
  \country{UK}
}
\email{2565278@dundee.ac.uk}

\author{Michael Crabb}
\orcid{0000-0002-9563-0691}
\affiliation{%
  \institution{School of Science and Engineering\\University of Dundee}
  \city{Dundee}
  \country{UK}}
\email{m.z.crabb@dundee.ac.uk}







\renewcommand{\shortauthors}{You et al.}

\begin{abstract}
The consumption of subtitles via TVs, laptops and smartphones has the potential to marginalize people based on their complex accessibility needs. The current one-size-fits-all approach to this accessibility aid is no longer fit for purpose and work is required to look at how it can be adapted to be personalised for individual users based on individual context, content, and consumption habits.
People with Aphasia, for example, encounter significant challenges in understanding subtitle texts.

We see our work as a call to action for more inclusive practices, focusing on how the thoughts and opinions of people with aphasia can be included in media research.
Our work investigates how to develop future media solutions for people with aphasia to create a more inclusive media viewing environment.
We believe the key to this is appropriate prototyping tools and methods to allow equitable inclusion in the system design process.

\end{abstract}



\begin{teaserfigure}
  \centering
  \includegraphics[height=7cm, width=14cm]{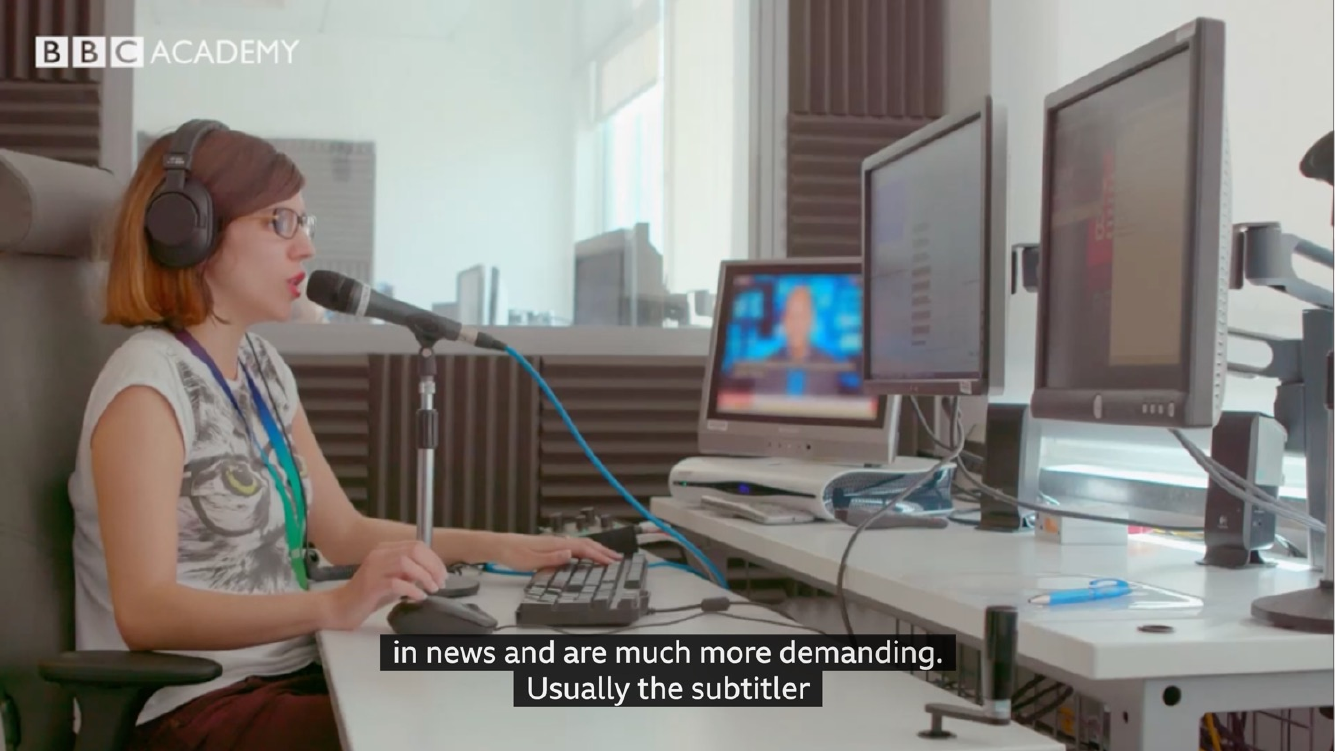}
  \caption{Traditional subtitle display on BBC.}
  \Description{Figure shows how traditional subtitle is displayed on BBC media platform.}
  \label{fig:teaser}
\end{teaserfigure}


\maketitle

\section{Current Realities: State-of-the-art subtitles accessibility research}
Subtitles/Closed Captions\footnote{Closed captions (CC) also provide a text description of sound effects. Most streaming sites only have the option for ‘English[CC]’ for English subtitles and therefore in this work, we collectively refer to both as subtitles.} textualise the film/television's spoken content~\cite{SUBTITLES_MEANING} and are traditionally displayed as white text with a black background at the bottom centre of the screen~\cite{BBC_GUIDELINE}, as suggested by BBC guidelines with an example in \autoref{fig:teaser}.
However, these are not equally accessible to everyone due to the one-size-fits-all approach to subtitle display and the temporal nature of subtitles being directly attached to time-stamped points within media.

The presentation of subtitles is heavily researched with work taking place over the last three decades focusing on improving this for all users.
In 1995, Brett~\cite{brett1995multimedia} used on-click explanations for complex subtitle words to enhance people's listening skills.
The animated subtitles that conveyed speakers' emotions were designed by Rashid et al.~\cite{rashid2006expressing} around ten years later.
External techniques and alternative modalities have started being considered in the 2020s, such as using people's eye gaze to gauge the subtitle placement~\cite{kurzhals2020view} and translating subtitle audio into vibrations that can be felt on the wrists of deaf and hard of hearing (DHH) people~\cite{wang2023haptic}.

Additional considerations have to be taken when creating subtitles for immersive interfaces.
In recent years, BBC Research \& Development proposed four subtitle behaviours in the virtual reality space~\cite{brown2017subtitles} and determined \textit{moving-with-head} subtitles~\cite{brown2018exploring} with 12.5 degrees below the eye-line~\cite{rothe2018positioning} as the optimal location from user studies.

\section{Future Perspectives: Developing accessible subtitled media}
Inspired by the current state of subtitles accessibility research, we explore future perspectives on accessible subtitled media.
This section addresses the population our research targets, and strategies to enhance user experience and comfort during media consumption.

\subsection{Community of focus}
Aphasia is a language and communication disorder onset by stroke that may impair a person's understanding, speaking, reading, writing, and using numbers~\cite{sherratt2024people}.
This stands for at least 350,000 people in the UK~\cite{APHASIA_STATISTICS}, facing challenges engaging in daily activities at home and workplace~\cite{dalemans2008description, parr2001psychosocial}, which have been addressed by researchers from several angles in recent years.
For instance, augmentative and alternative communication (AAC) strategies~\cite{millar1998augmentative} have been adopted to support their communication via wearable devices~\cite{curtis2023envisioning, curtis2024breaking}.

The digital content creation skills of people with aphasia have also been leveraged in terms of rearranging words to form new sentences~\cite{neate2019empowering}, making comic strips~\cite{tamburro2020accessible} and drawing~\cite{neate2020painting}.
While prior works have contributed valuable insights into several accessibility challenges people with aphasia encounter, enabling access to subtitled media remains necessary.
Moreover, an increase in traction of accessibility research within the subtitle space has been identified~\cite{nevsky2023accessibility}, where none of them focuses on the community of aphasia.
Given the growing significance of inclusive digital access and the complexity of subtitle display, it is crucial to understand the challenges of subtitle adaptation that meet their requirements.

\subsection{Design Thinking and Co-design}
To develop tailored solutions that improve engagement and understanding of subtitled media for people with aphasia, a combination of \textbf{Double Diamond Design Thinking}~\cite{DOUBLE_DIAMOND} and \textbf{Co-design}~\cite{sanders2008co} approaches have been proposed.
In our work, we plan to involve end users throughout the design, development, and evaluation of the accessibility interventions.
This strategy has gained attention in recent years, where researchers have been focusing on addressing accessibility challenges faced by older people~\cite{lindsay2012engaging, villena2014accessible} and disabled communities such as visual impairments~\cite{oliveira2017promoting, villena2014web}, hearing loss~\cite{villena2014web}, people with aphasia~\cite{curtis2023watch, nevsky2024lights} and dementia~\cite{lindsay2012empathy}.

To ensure that we develop appropriate digital artefacts, we need to properly understand the problem we are trying to solve by attempting to gain insights regarding the current challenges they face when consuming subtitles.
This can be achieved by conducting exploratory interviews with speech and language therapists (SLTs)~\cite{SLT} and surveying people with aphasia with appropriately structured closed questions~\cite{baunstrup2013elderly}, in which the findings can be synthesized to design guidelines of subtitle adaptation interventions to cater to the needs of people with aphasia.
The design guidance pinpoints the "rightness" of a solution, and the implementation will begin afterwards.

In our work, we plan to use eye-tracking devices\footnote{\href{https://www.tobii.com/products/eye-trackers}{https://www.tobii.com/products/eye-trackers}}~\cite{thaqi2024sara} to monitor users' current screen focus and subsequently use this to understand the overall effectiveness of our created system.
A horizontal prototyping~\cite{bacon2012prototyping} strategy will be considered, as it is likely that our solution will contain many independent/loosely dependent features that will be developed iteratively.
We will evaluate our prototype once most of the system features are functional.
This involves gathering a mixture of quantitative and qualitative data, highlighting the challenges and opportunities in developing systems that assist people with aphasia in consuming subtitled media content.
Our intervention will be compared pairwise with other existing accessibility interventions (i.e., users select one of the two or neutral)~\cite{tundik2020low}, complemented by semi-structured interviews~\cite{crabb2015online} that allow them to elaborate on their responses.

\section{Conclusion}
In our work we aim to improve the ability of people with aphasia to consume subtitled media content.
We contribute to HCI research in three main aspects~\cite{wobbrock2016research}.
First, insights from surveys and interviews with SLTs enable us to draft user requirements and design recommendations for the subtitle adaptation intervention, which correspond to a combination of empirical and theoretical contributions.
Our second contribution is in the development and implementation of a high-fidelity prototype that personalizes and customizes the subtitled media for people with aphasia.
Our final empirical contribution will be made from findings based on data collected from user studies that capture the experience of people with aphasia when interacting with our accessibility intervention.


\bibliographystyle{ACM-Reference-Format}
\bibliography{sample-base}


\begin{thebibliography}{36}


\ifx \showCODEN    \undefined \def \showCODEN     #1{\unskip}     \fi
\ifx \showISBNx    \undefined \def \showISBNx     #1{\unskip}     \fi
\ifx \showISBNxiii \undefined \def \showISBNxiii  #1{\unskip}     \fi
\ifx \showISSN     \undefined \def \showISSN      #1{\unskip}     \fi
\ifx \showLCCN     \undefined \def \showLCCN      #1{\unskip}     \fi
\ifx \shownote     \undefined \def \shownote      #1{#1}          \fi
\ifx \showarticletitle \undefined \def \showarticletitle #1{#1}   \fi
\ifx \showURL      \undefined \def \showURL       {\relax}        \fi
\providecommand\bibfield[2]{#2}
\providecommand\bibinfo[2]{#2}
\providecommand\natexlab[1]{#1}
\providecommand\showeprint[2][]{arXiv:#2}

\bibitem[Association(2025)]%
        {APHASIA_STATISTICS}
\bibfield{author}{\bibinfo{person}{Stroke Association}.} \bibinfo{year}{2025}\natexlab{}.
\newblock \bibinfo{booktitle}{\emph{Aphasia awareness and the communications access symbol}}.
\newblock
\urldef\tempurl%
\url{https://www.stroke.org.uk/stroke/effects/aphasia/aphasia-awareness#:~:text=More%20than%20350%2C000%20people%20in,disorder%20of%20language%20and%20communication.}
\showURL{%
Retrieved February 07, 2025 from \tempurl}


\bibitem[Bacon et~al\mbox{.}(2012)]%
        {bacon2012prototyping}
\bibfield{author}{\bibinfo{person}{Philip Bacon}, \bibinfo{person}{Reinhard Budde}, \bibinfo{person}{Karlheinz Kautz}, \bibinfo{person}{Karin Kuhlenkamp}, {and} \bibinfo{person}{Heinz Z{\"u}llighoven}.} \bibinfo{year}{2012}\natexlab{}.
\newblock \bibinfo{booktitle}{\emph{Prototyping: An Approach to Evolutionary System Development} (\bibinfo{edition}{1st.} ed.)}.
\newblock \bibinfo{publisher}{Springer-Verlag}, \bibinfo{address}{Berlin, Heidelberg}.
\newblock


\bibitem[Baunstrup and Larsen(2013)]%
        {baunstrup2013elderly}
\bibfield{author}{\bibinfo{person}{Mai Baunstrup} {and} \bibinfo{person}{Lars~Bo Larsen}.} \bibinfo{year}{2013}\natexlab{}.
\newblock \showarticletitle{Elderly’s barriers and requirements for interactive TV}. In \bibinfo{booktitle}{\emph{Universal Access in Human-Computer Interaction. User and Context Diversity: 7th International Conference, UAHCI 2013, Held as Part of HCI International 2013, Las Vegas, NV, USA, July 21-26, 2013, Proceedings, Part II 7}} \emph{(\bibinfo{series}{UAHCI '13})}. \bibinfo{publisher}{Springer}, \bibinfo{address}{Berlin/Heidelberg, Germany}, \bibinfo{pages}{13--22}.
\newblock
\href{https://doi.org/10.1007/978-3-642-39191-0_2}{doi:\nolinkurl{10.1007/978-3-642-39191-0_2}}


\bibitem[BBC(2024)]%
        {BBC_GUIDELINE}
\bibfield{author}{\bibinfo{person}{BBC}.} \bibinfo{year}{2024}\natexlab{}.
\newblock \bibinfo{booktitle}{\emph{Subtitle Guidelines}}.
\newblock
\urldef\tempurl%
\url{https://www.bbc.co.uk/accessibility/forproducts/guides/subtitles/}
\showURL{%
Retrieved February 06, 2025 from \tempurl}


\bibitem[Brett(1995)]%
        {brett1995multimedia}
\bibfield{author}{\bibinfo{person}{Paul Brett}.} \bibinfo{year}{1995}\natexlab{}.
\newblock \showarticletitle{Multimedia for listening comprehension: The design of a multimedia-based resource for developing listening skills}.
\newblock \bibinfo{journal}{\emph{System}} \bibinfo{volume}{23}, \bibinfo{number}{1} (\bibinfo{date}{Feb.} \bibinfo{year}{1995}), \bibinfo{pages}{77--85}.
\newblock
\href{https://doi.org/10.1016/0346-251X(94)00054-A}{doi:\nolinkurl{10.1016/0346-251X(94)00054-A}}


\bibitem[Brown et~al\mbox{.}(2017)]%
        {brown2017subtitles}
\bibfield{author}{\bibinfo{person}{Andy Brown}, \bibinfo{person}{Jayson Turner}, \bibinfo{person}{Jake Patterson}, \bibinfo{person}{Anastasia Schmitz}, \bibinfo{person}{Mike Armstrong}, {and} \bibinfo{person}{Maxine Glancy}.} \bibinfo{year}{2017}\natexlab{}.
\newblock \showarticletitle{Subtitles in 360-degree Video}. In \bibinfo{booktitle}{\emph{Adjunct Publication of the 2017 ACM International Conference on Interactive Experiences for TV and Online Video}} \emph{(\bibinfo{series}{TVX '17})}. \bibinfo{publisher}{Association for Computing Machinery}, \bibinfo{address}{New York, NY, USA}, \bibinfo{pages}{5--6}.
\newblock
\href{https://doi.org/10.1145/3084289.3089915}{doi:\nolinkurl{10.1145/3084289.3089915}}


\bibitem[Brown et~al\mbox{.}(2018)]%
        {brown2018exploring}
\bibfield{author}{\bibinfo{person}{Andy Brown}, \bibinfo{person}{Jayson Turner}, \bibinfo{person}{Jake Patterson}, \bibinfo{person}{Anastasia Schmitz}, \bibinfo{person}{Mike Armstrong}, {and} \bibinfo{person}{Maxine Glancy}.} \bibinfo{year}{2018}\natexlab{}.
\newblock \bibinfo{booktitle}{\emph{Exploring subtitle behaviour for 360 video}}.
\newblock
\urldef\tempurl%
\url{https://downloads.bbc.co.uk/rd/pubs/whp/whp-pdf-files/WHP330.pdf}
\showURL{%
Retrieved February 07, 2025 from \tempurl}


\bibitem[Crabb et~al\mbox{.}(2015)]%
        {crabb2015online}
\bibfield{author}{\bibinfo{person}{Michael Crabb}, \bibinfo{person}{Rhianne Jones}, \bibinfo{person}{Mike Armstrong}, {and} \bibinfo{person}{Chris~J Hughes}.} \bibinfo{year}{2015}\natexlab{}.
\newblock \showarticletitle{Online news videos: the UX of subtitle position}. In \bibinfo{booktitle}{\emph{Proceedings of the 17th international ACM SIGACCESS conference on Computers \& accessibility}} \emph{(\bibinfo{series}{ASSETS '15})}. \bibinfo{publisher}{Association for Computing Machinery}, \bibinfo{address}{New York, NY, USA}, \bibinfo{pages}{218}.
\newblock
\href{https://doi.org/10.1145/2700648.2809866}{doi:\nolinkurl{10.1145/2700648.2809866}}


\bibitem[Curtis et~al\mbox{.}(2024)]%
        {curtis2024breaking}
\bibfield{author}{\bibinfo{person}{Humphrey Curtis}, \bibinfo{person}{Ying~Hei Lau}, {and} \bibinfo{person}{Timothy Neate}.} \bibinfo{year}{2024}\natexlab{}.
\newblock \showarticletitle{Breaking Badge: Augmenting Communication with Wearable AAC Smartbadges and Displays}. In \bibinfo{booktitle}{\emph{Proceedings of the 2024 CHI Conference on Human Factors in Computing Systems}} \emph{(\bibinfo{series}{CHI '24})}. \bibinfo{publisher}{Association for Computing Machinery}, \bibinfo{address}{New York, NY, USA}, \bibinfo{pages}{2}.
\newblock
\href{https://doi.org/10.1145/3613904.3642327}{doi:\nolinkurl{10.1145/3613904.3642327}}


\bibitem[Curtis and Neate(2023)]%
        {curtis2023watch}
\bibfield{author}{\bibinfo{person}{Humphrey Curtis} {and} \bibinfo{person}{Timothy Neate}.} \bibinfo{year}{2023}\natexlab{}.
\newblock \showarticletitle{Watch Your Language: Using Smartwatches To Support Communication}. In \bibinfo{booktitle}{\emph{Proceedings of the 25th International ACM SIGACCESS Conference on Computers and Accessibility}} \emph{(\bibinfo{series}{ASSETS '23})}. \bibinfo{publisher}{Association for Computing Machinery}, \bibinfo{address}{New York, NY, USA}, \bibinfo{pages}{1--21}.
\newblock
\href{https://doi.org/10.1145/3597638.3608379}{doi:\nolinkurl{10.1145/3597638.3608379}}


\bibitem[Curtis et~al\mbox{.}(2023)]%
        {curtis2023envisioning}
\bibfield{author}{\bibinfo{person}{Humphrey Curtis}, \bibinfo{person}{Zihao You}, \bibinfo{person}{William Deary}, \bibinfo{person}{Miruna-Ioana Tudoreanu}, {and} \bibinfo{person}{Timothy Neate}.} \bibinfo{year}{2023}\natexlab{}.
\newblock \showarticletitle{Envisioning the (in) visibility of discreet and wearable AAC devices}. In \bibinfo{booktitle}{\emph{Proceedings of the 2023 CHI Conference on Human Factors in Computing Systems}} \emph{(\bibinfo{series}{CHI '23})}. \bibinfo{publisher}{Association for Computing Machinery}, \bibinfo{address}{New York, NY, USA}, \bibinfo{pages}{2}.
\newblock
\href{https://doi.org/10.1145/3544548.3580936}{doi:\nolinkurl{10.1145/3544548.3580936}}


\bibitem[Dalemans et~al\mbox{.}(2008)]%
        {dalemans2008description}
\bibfield{author}{\bibinfo{person}{Ruth J.~P. Dalemans}, \bibinfo{person}{Luc P.~De Witte}, \bibinfo{person}{Derick~T. Wade}, {and} \bibinfo{person}{Wim J. A.~Van den Heuvel}.} \bibinfo{year}{2008}\natexlab{}.
\newblock \showarticletitle{A description of social participation in working-age persons with aphasia: A review of the literature}.
\newblock \bibinfo{journal}{\emph{Aphasiology}} \bibinfo{volume}{22}, \bibinfo{number}{10} (\bibinfo{date}{Oct.} \bibinfo{year}{2008}), \bibinfo{pages}{1072}.
\newblock
\href{https://doi.org/10.1080/02687030701632179}{doi:\nolinkurl{10.1080/02687030701632179}}


\bibitem[Humble(2025)]%
        {DOUBLE_DIAMOND}
\bibfield{author}{\bibinfo{person}{Jeff Humble}.} \bibinfo{year}{2025}\natexlab{}.
\newblock \bibinfo{booktitle}{\emph{What is the Double Diamond Design Process?}}
\newblock
\urldef\tempurl%
\url{https://www.thefountaininstitute.com/blog/what-is-the-double-diamond-design-process}
\showURL{%
Retrieved February 10, 2025 from \tempurl}


\bibitem[Kurzhals et~al\mbox{.}(2020)]%
        {kurzhals2020view}
\bibfield{author}{\bibinfo{person}{Kuno Kurzhals}, \bibinfo{person}{Fabian Göbel}, \bibinfo{person}{Katrin Angerbauer}, \bibinfo{person}{Michael Sedlmair}, {and} \bibinfo{person}{Martin Raubal}.} \bibinfo{year}{2020}\natexlab{}.
\newblock \showarticletitle{A view on the viewer: Gaze-adaptive captions for videos}. In \bibinfo{booktitle}{\emph{Proceedings of the 2020 CHI Conference on Human Factors in Computing Systems}} \emph{(\bibinfo{series}{CHI '20})}. \bibinfo{publisher}{Association for Computing Machinery}, \bibinfo{address}{New York, NY, USA}, \bibinfo{pages}{1--12}.
\newblock
\href{https://doi.org/10.1145/3313831.3376266}{doi:\nolinkurl{10.1145/3313831.3376266}}


\bibitem[Lindsay et~al\mbox{.}(2012a)]%
        {lindsay2012empathy}
\bibfield{author}{\bibinfo{person}{Stephen Lindsay}, \bibinfo{person}{Katie Brittain}, \bibinfo{person}{Daniel Jackson}, \bibinfo{person}{Cassim Ladha}, \bibinfo{person}{Karim Ladha}, {and} \bibinfo{person}{Patrick Olivier}.} \bibinfo{year}{2012}\natexlab{a}.
\newblock \showarticletitle{Empathy, participatory design and people with dementia}. In \bibinfo{booktitle}{\emph{Proceedings of the SIGCHI conference on Human factors in computing systems}} \emph{(\bibinfo{series}{CHI '12})}. \bibinfo{publisher}{Association for Computing Machinery}, \bibinfo{address}{New York, NY, USA}, \bibinfo{pages}{521--530}.
\newblock
\href{https://doi.org/10.1145/2207676.2207749}{doi:\nolinkurl{10.1145/2207676.2207749}}


\bibitem[Lindsay et~al\mbox{.}(2012b)]%
        {lindsay2012engaging}
\bibfield{author}{\bibinfo{person}{Stephen Lindsay}, \bibinfo{person}{Daniel Jackson}, \bibinfo{person}{Guy Schofield}, {and} \bibinfo{person}{Patrick Olivier}.} \bibinfo{year}{2012}\natexlab{b}.
\newblock \showarticletitle{Engaging older people using participatory design}. In \bibinfo{booktitle}{\emph{Proceedings of the SIGCHI conference on human factors in computing systems}} \emph{(\bibinfo{series}{CHI '12})}. \bibinfo{publisher}{Association for Computing Machinery}, \bibinfo{address}{New York, NY, USA}, \bibinfo{pages}{1199--1208}.
\newblock
\href{https://doi.org/10.1145/2207676.2208570}{doi:\nolinkurl{10.1145/2207676.2208570}}


\bibitem[Millar and Scott(1998)]%
        {millar1998augmentative}
\bibfield{author}{\bibinfo{person}{Sally Millar} {and} \bibinfo{person}{Janet Scott}.} \bibinfo{year}{1998}\natexlab{}.
\newblock \bibinfo{booktitle}{\emph{What is augmentative and alternative communication? An introduction}}.
\newblock
\urldef\tempurl%
\url{https://citeseerx.ist.psu.edu/document?repid=rep1&type=pdf&doi=52f220ecc9ac08b8d6d20a82f73fb4efa2c8aa09}
\showURL{%
Retrieved February 07, 2025 from \tempurl}


\bibitem[Neate et~al\mbox{.}(2020)]%
        {neate2020painting}
\bibfield{author}{\bibinfo{person}{Timothy Neate}, \bibinfo{person}{Abi Roper}, {and} \bibinfo{person}{Stephanie Wilson}.} \bibinfo{year}{2020}\natexlab{}.
\newblock \showarticletitle{Painting a Picture of Accessible Digital Art}. In \bibinfo{booktitle}{\emph{Proceedings of the 22nd International ACM SIGACCESS Conference on Computers and Accessibility}} \emph{(\bibinfo{series}{ASSETS '20})}. \bibinfo{publisher}{Association for Computing Machinery}, \bibinfo{address}{New York, NY, USA}, \bibinfo{pages}{1--3}.
\newblock
\href{https://doi.org/10.1145/3373625.3418019}{doi:\nolinkurl{10.1145/3373625.3418019}}


\bibitem[Neate et~al\mbox{.}(2019)]%
        {neate2019empowering}
\bibfield{author}{\bibinfo{person}{Timothy Neate}, \bibinfo{person}{Abi Roper}, \bibinfo{person}{Stephanie Wilson}, {and} \bibinfo{person}{Jane Marshall}.} \bibinfo{year}{2019}\natexlab{}.
\newblock \showarticletitle{Empowering expression for users with aphasia through constrained creativity}. In \bibinfo{booktitle}{\emph{Proceedings of the 2019 CHI conference on human factors in computing systems}} \emph{(\bibinfo{series}{CHI '19})}. \bibinfo{publisher}{Association for Computing Machinery}, \bibinfo{address}{New York, NY, USA}, \bibinfo{pages}{6}.
\newblock
\href{https://doi.org/10.1145/3290605.3300615}{doi:\nolinkurl{10.1145/3290605.3300615}}


\bibitem[Nevsky et~al\mbox{.}(2024)]%
        {nevsky2024lights}
\bibfield{author}{\bibinfo{person}{Alexandre Nevsky}, \bibinfo{person}{Timothy Neate}, \bibinfo{person}{Elena Simperl}, {and} \bibinfo{person}{Madeline~N Cruice}.} \bibinfo{year}{2024}\natexlab{}.
\newblock \showarticletitle{Lights, Camera, Access: A Closeup on Audiovisual Media Accessibility and Aphasia}. In \bibinfo{booktitle}{\emph{Proceedings of the 2024 CHI Conference on Human Factors in Computing Systems}} \emph{(\bibinfo{series}{CHI '24})}. \bibinfo{publisher}{Association for Computing Machinery}, \bibinfo{address}{New York, NY, USA}, \bibinfo{pages}{1--17}.
\newblock
\href{https://doi.org/10.1145/3613904.3641893}{doi:\nolinkurl{10.1145/3613904.3641893}}


\bibitem[Nevsky et~al\mbox{.}(2023)]%
        {nevsky2023accessibility}
\bibfield{author}{\bibinfo{person}{Alexandre Nevsky}, \bibinfo{person}{Timothy Neate}, \bibinfo{person}{Elena Simperl}, {and} \bibinfo{person}{Radu-Daniel Vatavu}.} \bibinfo{year}{2023}\natexlab{}.
\newblock \showarticletitle{Accessibility Research in Digital Audiovisual Media: What Has Been Achieved and What Should Be Done Next?}. In \bibinfo{booktitle}{\emph{Proceedings of the 2023 ACM International Conference on Interactive Media Experiences}} \emph{(\bibinfo{series}{IMX '23})}. \bibinfo{publisher}{Association for Computing Machinery}, \bibinfo{address}{New York, NY, USA}, \bibinfo{pages}{94--114}.
\newblock
\href{https://doi.org/10.1145/3573381.3596159}{doi:\nolinkurl{10.1145/3573381.3596159}}


\bibitem[NHS(2025)]%
        {SLT}
\bibfield{author}{\bibinfo{person}{NHS}.} \bibinfo{year}{2025}\natexlab{}.
\newblock \bibinfo{booktitle}{\emph{Speech and language therapist}}.
\newblock
\urldef\tempurl%
\url{https://www.healthcareers.nhs.uk/explore-roles/allied-health-professionals/roles-allied-health-professions/speech-and-language-therapist}
\showURL{%
Retrieved February 10, 2025 from \tempurl}


\bibitem[Oliveira et~al\mbox{.}(2017)]%
        {oliveira2017promoting}
\bibfield{author}{\bibinfo{person}{Rita Oliveira}, \bibinfo{person}{Jorge~Ferraz de Abreu}, {and} \bibinfo{person}{Ana~Margarida Almeida}.} \bibinfo{year}{2017}\natexlab{}.
\newblock \showarticletitle{Promoting interactive television (iTV) accessibility: an adapted service for users with visual impairments}.
\newblock \bibinfo{journal}{\emph{Universal Access in the Information Society}}  \bibinfo{volume}{16} (\bibinfo{date}{Aug.} \bibinfo{year}{2017}), \bibinfo{pages}{533--544}.
\newblock
\href{https://doi.org/10.1007/s10209-016-0482-z}{doi:\nolinkurl{10.1007/s10209-016-0482-z}}


\bibitem[Parr(2001)]%
        {parr2001psychosocial}
\bibfield{author}{\bibinfo{person}{Susie Parr}.} \bibinfo{year}{2001}\natexlab{}.
\newblock \showarticletitle{Psychosocial aspects of aphasia: whose perspectives?}
\newblock \bibinfo{journal}{\emph{Folia phoniatrica et logopaedica}} \bibinfo{volume}{53}, \bibinfo{number}{5} (\bibinfo{date}{July} \bibinfo{year}{2001}), \bibinfo{pages}{274}.
\newblock
\href{https://doi.org/10.1159/000052681}{doi:\nolinkurl{10.1159/000052681}}


\bibitem[Rashid et~al\mbox{.}(2006)]%
        {rashid2006expressing}
\bibfield{author}{\bibinfo{person}{Raisa Rashid}, \bibinfo{person}{Jonathan Aitken}, {and} \bibinfo{person}{Deborah~I. Fels}.} \bibinfo{year}{2006}\natexlab{}.
\newblock \showarticletitle{Expressing emotions using animated text captions}. In \bibinfo{booktitle}{\emph{Computers Helping People with Special Needs: 10th International Conference, ICCHP 2006, Linz, Austria, July 11-13, 2006. Proceedings 10}} \emph{(\bibinfo{series}{ICCHP '06})}. \bibinfo{publisher}{Springer}, \bibinfo{address}{Berlin, Heidelberg}, \bibinfo{pages}{24--31}.
\newblock
\href{https://doi.org/10.1007/11788713_5}{doi:\nolinkurl{10.1007/11788713_5}}


\bibitem[Rothe et~al\mbox{.}(2018)]%
        {rothe2018positioning}
\bibfield{author}{\bibinfo{person}{Sylvia Rothe}, \bibinfo{person}{Kim Tran}, {and} \bibinfo{person}{Heinrich Hussmann}.} \bibinfo{year}{2018}\natexlab{}.
\newblock \showarticletitle{Positioning of Subtitles in Cinematic Virtual Reality.}. In \bibinfo{booktitle}{\emph{International Conference on Artificial Reality and Telexistence and Eurographics Symposium on Virtual Environments}} \emph{(\bibinfo{series}{ICAT-EGVE '18})}. \bibinfo{publisher}{The Eurographics Association}, \bibinfo{address}{Eindhoven, Netherlands}, \bibinfo{pages}{3}.
\newblock
\href{https://doi.org/10.2312/egve.20181307}{doi:\nolinkurl{10.2312/egve.20181307}}


\bibitem[Sanders and Stappers(2008)]%
        {sanders2008co}
\bibfield{author}{\bibinfo{person}{Elizabeth B.-N. Sanders} {and} \bibinfo{person}{Pieter~Jan Stappers}.} \bibinfo{year}{2008}\natexlab{}.
\newblock \showarticletitle{Co-creation and the new landscapes of design}.
\newblock \bibinfo{journal}{\emph{Co-design}} \bibinfo{volume}{4}, \bibinfo{number}{1} (\bibinfo{date}{March} \bibinfo{year}{2008}), \bibinfo{pages}{5--18}.
\newblock
\href{https://doi.org/10.1080/15710880701875068}{doi:\nolinkurl{10.1080/15710880701875068}}


\bibitem[Sherratt(2024)]%
        {sherratt2024people}
\bibfield{author}{\bibinfo{person}{Sue Sherratt}.} \bibinfo{year}{2024}\natexlab{}.
\newblock \showarticletitle{People with aphasia living alone: A scoping review}.
\newblock \bibinfo{journal}{\emph{Aphasiology}} \bibinfo{volume}{38}, \bibinfo{number}{4} (\bibinfo{date}{April} \bibinfo{year}{2024}), \bibinfo{pages}{713}.
\newblock
\href{https://doi.org/10.1080/02687038.2023.2227403}{doi:\nolinkurl{10.1080/02687038.2023.2227403}}


\bibitem[Tamburro et~al\mbox{.}(2020)]%
        {tamburro2020accessible}
\bibfield{author}{\bibinfo{person}{Carla Tamburro}, \bibinfo{person}{Timothy Neate}, \bibinfo{person}{Abi Roper}, {and} \bibinfo{person}{Stephanie Wilson}.} \bibinfo{year}{2020}\natexlab{}.
\newblock \showarticletitle{Accessible creativity with a comic spin}. In \bibinfo{booktitle}{\emph{Proceedings of the 22nd International ACM SIGACCESS Conference on Computers and Accessibility}} \emph{(\bibinfo{series}{ASSETS '20})}. \bibinfo{publisher}{Association for Computing Machinery}, \bibinfo{address}{New York, NY, USA}, \bibinfo{pages}{3--5}.
\newblock
\href{https://doi.org/10.1145/3373625.3417012}{doi:\nolinkurl{10.1145/3373625.3417012}}


\bibitem[Thaqi et~al\mbox{.}(2024)]%
        {thaqi2024sara}
\bibfield{author}{\bibinfo{person}{Enkeleda Thaqi}, \bibinfo{person}{Mohamed~Omar Mantawy}, {and} \bibinfo{person}{Enkelejda Kasneci}.} \bibinfo{year}{2024}\natexlab{}.
\newblock \showarticletitle{SARA: Smart AI Reading Assistant for Reading Comprehension}. In \bibinfo{booktitle}{\emph{Proceedings of the 2024 Symposium on Eye Tracking Research and Applications}} \emph{(\bibinfo{series}{ETRA '24})}. \bibinfo{publisher}{Association for Computing Machinery}, \bibinfo{address}{New York, NY, USA}, \bibinfo{pages}{1--3}.
\newblock
\href{https://doi.org/10.1145/3649902.3655661}{doi:\nolinkurl{10.1145/3649902.3655661}}


\bibitem[Thesaurus(2025)]%
        {SUBTITLES_MEANING}
\bibfield{author}{\bibinfo{person}{Cambridge Advanced Learner's Dictionary~\& Thesaurus}.} \bibinfo{year}{2025}\natexlab{}.
\newblock \bibinfo{booktitle}{\emph{Meaning of subtitles in English}}.
\newblock
\urldef\tempurl%
\url{https://dictionary.cambridge.org/dictionary/english/subtitles}
\showURL{%
Retrieved February 06, 2025 from \tempurl}


\bibitem[T{\"u}ndik et~al\mbox{.}(2020)]%
        {tundik2020low}
\bibfield{author}{\bibinfo{person}{M{\'a}t{\'e}~{\'A}kos T{\"u}ndik}, \bibinfo{person}{Bal{\'a}zs Tarj{\'a}n}, {and} \bibinfo{person}{Gy{\"o}rgy Szasz{\'a}k}.} \bibinfo{year}{2020}\natexlab{}.
\newblock \showarticletitle{A low latency sequential model and its user-focused evaluation for automatic punctuation of ASR closed captions}.
\newblock \bibinfo{journal}{\emph{Computer Speech \& Language}}  \bibinfo{volume}{63} (\bibinfo{date}{Sept.} \bibinfo{year}{2020}), \bibinfo{pages}{15--17}.
\newblock
\href{https://doi.org/10.1016/j.csl.2020.101076}{doi:\nolinkurl{10.1016/j.csl.2020.101076}}


\bibitem[Villena et~al\mbox{.}(2014a)]%
        {villena2014accessible}
\bibfield{author}{\bibinfo{person}{Johana M~Rosas Villena}, \bibinfo{person}{Bruno~C Ramos}, \bibinfo{person}{Renata~PM Fortes}, {and} \bibinfo{person}{Rudinei Goularte}.} \bibinfo{year}{2014}\natexlab{a}.
\newblock \showarticletitle{An accessible video player for older people: Issues from a user test}.
\newblock \bibinfo{journal}{\emph{Procedia Computer Science}}  \bibinfo{volume}{27} (\bibinfo{date}{Jan.} \bibinfo{year}{2014}), \bibinfo{pages}{168--175}.
\newblock
\href{https://doi.org/10.1016/j.procs.2014.02.020}{doi:\nolinkurl{10.1016/j.procs.2014.02.020}}


\bibitem[Villena et~al\mbox{.}(2014b)]%
        {villena2014web}
\bibfield{author}{\bibinfo{person}{Johana Mar{\'\i}a~Rosas Villena}, \bibinfo{person}{Bruno~Costa Ramos}, \bibinfo{person}{Renata Pontin~M Fortes}, {and} \bibinfo{person}{Rudinei Goularte}.} \bibinfo{year}{2014}\natexlab{b}.
\newblock \showarticletitle{Web videos--concerns about accessibility based on user centered design}.
\newblock \bibinfo{journal}{\emph{Procedia Computer Science}}  \bibinfo{volume}{27} (\bibinfo{date}{Jan.} \bibinfo{year}{2014}), \bibinfo{pages}{481--490}.
\newblock
\href{https://doi.org/10.1016/j.procs.2014.02.052}{doi:\nolinkurl{10.1016/j.procs.2014.02.052}}


\bibitem[Wang et~al\mbox{.}(2023)]%
        {wang2023haptic}
\bibfield{author}{\bibinfo{person}{Yiwen Wang}, \bibinfo{person}{Ziming Li}, \bibinfo{person}{Pratheep~Kumar Chelladurai}, \bibinfo{person}{Wendy Dannels}, \bibinfo{person}{Tae Oh}, {and} \bibinfo{person}{Roshan~L Peiris}.} \bibinfo{year}{2023}\natexlab{}.
\newblock \showarticletitle{Haptic-Captioning: Using Audio-Haptic Interfaces to Enhance Speaker Indication in Real-Time Captions for Deaf and Hard-of-Hearing Viewers}. In \bibinfo{booktitle}{\emph{Proceedings of the 2023 CHI Conference on Human Factors in Computing Systems}} \emph{(\bibinfo{series}{CHI '23})}. \bibinfo{publisher}{Association for Computing Machinery}, \bibinfo{address}{New York, NY, USA}, \bibinfo{pages}{1--14}.
\newblock
\href{https://doi.org/10.1145/3544548.3581076}{doi:\nolinkurl{10.1145/3544548.3581076}}


\bibitem[Wobbrock and Kientz(2016)]%
        {wobbrock2016research}
\bibfield{author}{\bibinfo{person}{Jacob~O. Wobbrock} {and} \bibinfo{person}{Julie~A. Kientz}.} \bibinfo{year}{2016}\natexlab{}.
\newblock \showarticletitle{Research contributions in human-computer interaction}.
\newblock \bibinfo{journal}{\emph{interactions}} \bibinfo{volume}{23}, \bibinfo{number}{3} (\bibinfo{date}{April} \bibinfo{year}{2016}), \bibinfo{pages}{38--44}.
\newblock
\href{https://doi.org/10.1145/2907069}{doi:\nolinkurl{10.1145/2907069}}


\end{thebibliography}

\end{document}